\documentclass[prb,10pt,twocolumn,superscriptaddress]{revtex4-2}

\usepackage{graphicx}
\usepackage{dcolumn}
\usepackage{bm}
\usepackage{gensymb}

\usepackage[utf8]{inputenc}
\usepackage[T1]{fontenc}
\usepackage{mathptmx}
\begin{document}


\title{Antiferromagnetic Spin Orientation and Magnetic Domain Structure in Epitaxially Grown MnN Studied using Optical Second Harmonic Generation}

\author{Joongwon Lee}
\email{jl3755@cornell.edu}

\author{Zexuan Zhang}
 \email{zz523@cornell.edu}
\affiliation{School of Electrical and Computer Engineering, Cornell University, Ithaca NY 14853, USA}

\author{Huili (Grace) Xing}
\affiliation{School of Electrical and Computer Engineering, Cornell University, Ithaca NY 14853, USA}
\affiliation{Department of Materials Science and Engineering, Cornell University, Ithaca NY 14853, USA}

\author{Debdeep Jena}
\affiliation{School of Electrical and Computer Engineering, Cornell University, Ithaca NY 14853, USA}
\affiliation{Department of Materials Science and Engineering, Cornell University, Ithaca NY 14853, USA}

\author{Farhan Rana}
\affiliation{School of Electrical and Computer Engineering, Cornell University, Ithaca NY 14853, USA}

\date{\today}

\begin{abstract}
MnN is a centrosymmetric collinear antiferromagnet belonging to the transition metal nitride family with a high Neel temperature ($\sim$660 K), a low anisotropy field, and a large magnetic moment ($\sim$3.3 $\mu_{B}$) per Mn atom. Despite several recent experimental and theoretical studies, the spin symmetry (magnetic point group) and magnetic domain structure of the material remain unknown. In this work, we use optical second harmonic generation (SHG) to study the magnetic structure of thin epitaxially-grown single-crystal (001) MnN films. Our work shows that spin moments in MnN are tilted away from the [001] direction and the components of the spin moments in the (001) plane are aligned along one of the the two possible in-plane symmetry axes ([100] or [110]) resulting in a magnetic point group symmetry of 2/m1'. Our work rules out magnetic point group symmetries 4/mmm1' and mmm1' that have been previously discussed in the literature. Four different spin domains consistent with the 2/m1' magnetic point group symmetry are possible in MnN. A statistical model based on the observed variations in the polarization-dependent intensity of the second harmonic signal collected over large sample areas puts an upper bound of $\sim$0.65 $\mu$m on the mean domain size. Our results show that SHG can be used to probe the magnetic order in metallic antiferromagnets. This work is expected to contribute to the recent efforts in using antiferromagnets for spintronic applications.  
\end{abstract}

\maketitle

\section{Introduction}

\begin{figure}[t]
\includegraphics[width=0.95\columnwidth]{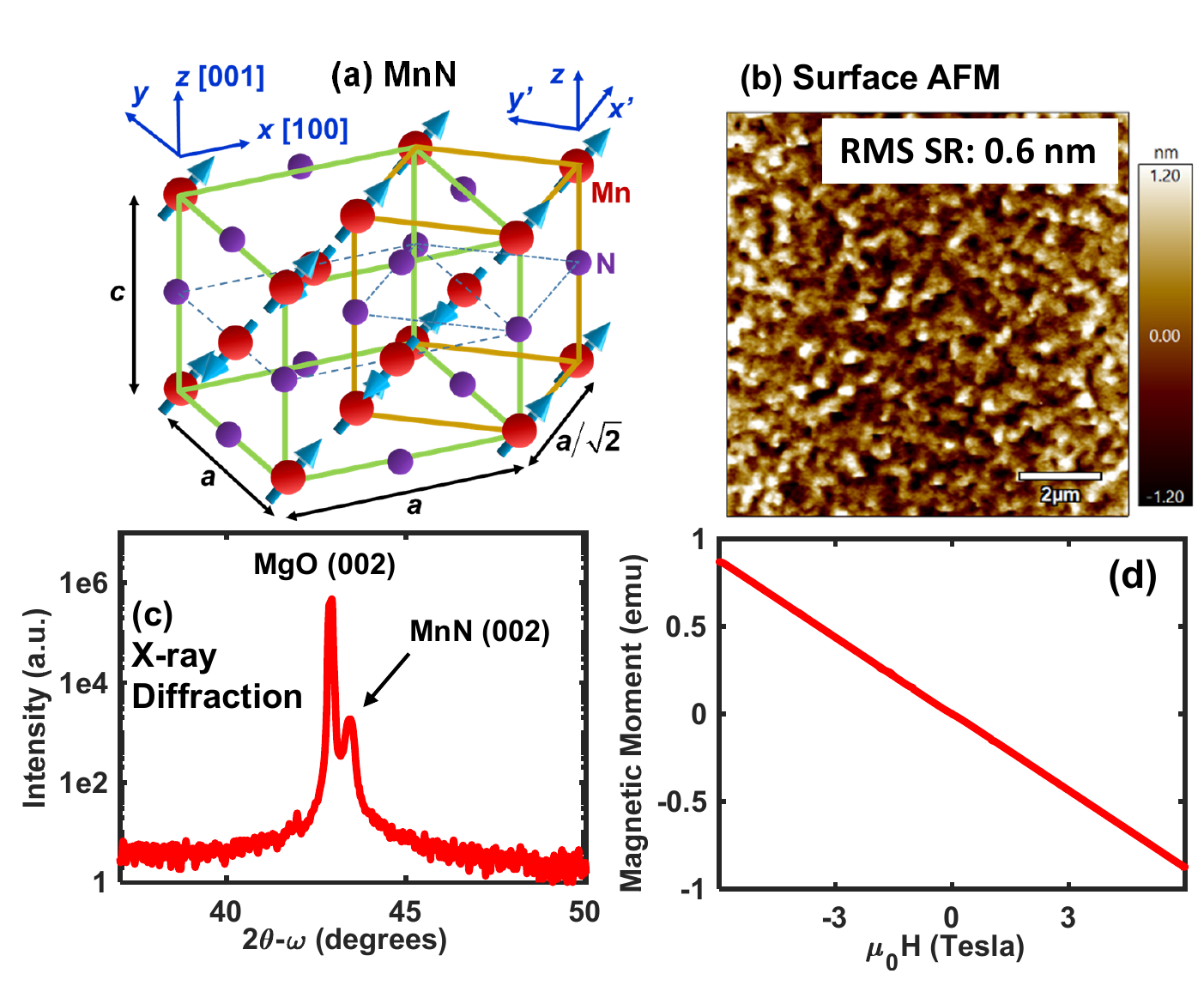}
\caption{\label{Growth} (a) Unit cells and magnetic ordering in MnN (2/m1' symmetry) are shown. Both the Fcc (unprimed axes) and Bcc (primed axes) unit cells of the tetragonal crystal are depicted. (b) AFM scan of the surface of epitaxially grown (001) MnN. RMS surface roughness (SR) is 0.6 nm. (c) X-ray diffraction pattern of epitaxially grown MnN. (d) Room temperature out-of-plane magnetic hysteresis study of a 500 nm thick MnN sample (grown on (001) MgO).}
\end{figure}

Antiferromagnetic (AF) materials have drawn considerable attention recently due to their potential for applications in spintronics~\cite{Rev1,Rev2}. The transition metal nitrides are particularly interesting in this context because of the extensive range of electronic and magnetic properties they display~\cite{TMN1985}. Mn-containing nitrides are especially interesting, for they form a variety of stoichiometric phases with Mn ions in various valence states. Among these phases one finds both antiferromagnets and ferrimagnets~\cite{Suzuki2001, Leineweber2000,Kleinman2003,Lamb2003,Lamb2005,Zunger2008}. The diversity of magnetic orders, combined with the ability to epitaxially grow high-quality single-crystal films~\cite{Jena2019}, makes Mn nitrides attractive for spintronics~\cite{Meinert2017,Dunz2020}.

Spin-torque switching of Neel order has been reported for several metallic and insulating AF materials~\cite{Olejnik2017,Bodnar2018,Chen2018,Moriyama2018,Cheng2020}. Recently, electrical spin-torque switching has also been reported for MnN stacked with spin-Hall metal Pt~\cite{Dunz2020}. In many of these experiments, electrical read-out schemes to detect the Neel order, and its switching, are weak. Various optical, X-ray, and thermal schemes have been proposed and used to investigate the Neel order ~\cite{Gray2019,Meer2020}. Linear optical techniques, such as Faraday/Kerr rotation or linear/circular dichroism, don't generally work in fully compensated AF materials other than in special cases where DC electric (or magnetic) fields can be applied to generate a linear optical response in AF materials lacking inversion (or time-reversal) symmetry~\cite{Khar1987}, or where the magnetic order in the AF material is strongly coupled to the crystal structure and lattice strain~\cite{Meer2020}. Nonlinear optical techniques, and SHG in particular, have been shown to be sensitive to the Neel order and magnetic domain structure in AF materials~\cite{Fiebig2005,Sanger2006} as a result of the magnetic dipole and electric quadrupole transitions in these materials. In this work, we use SHG to determine the magnetic point group symmetry and magnetic domain structure of epitaxially-grown $\theta$-phase MnN films.

\begin{figure}
\includegraphics[width=1.0\columnwidth]{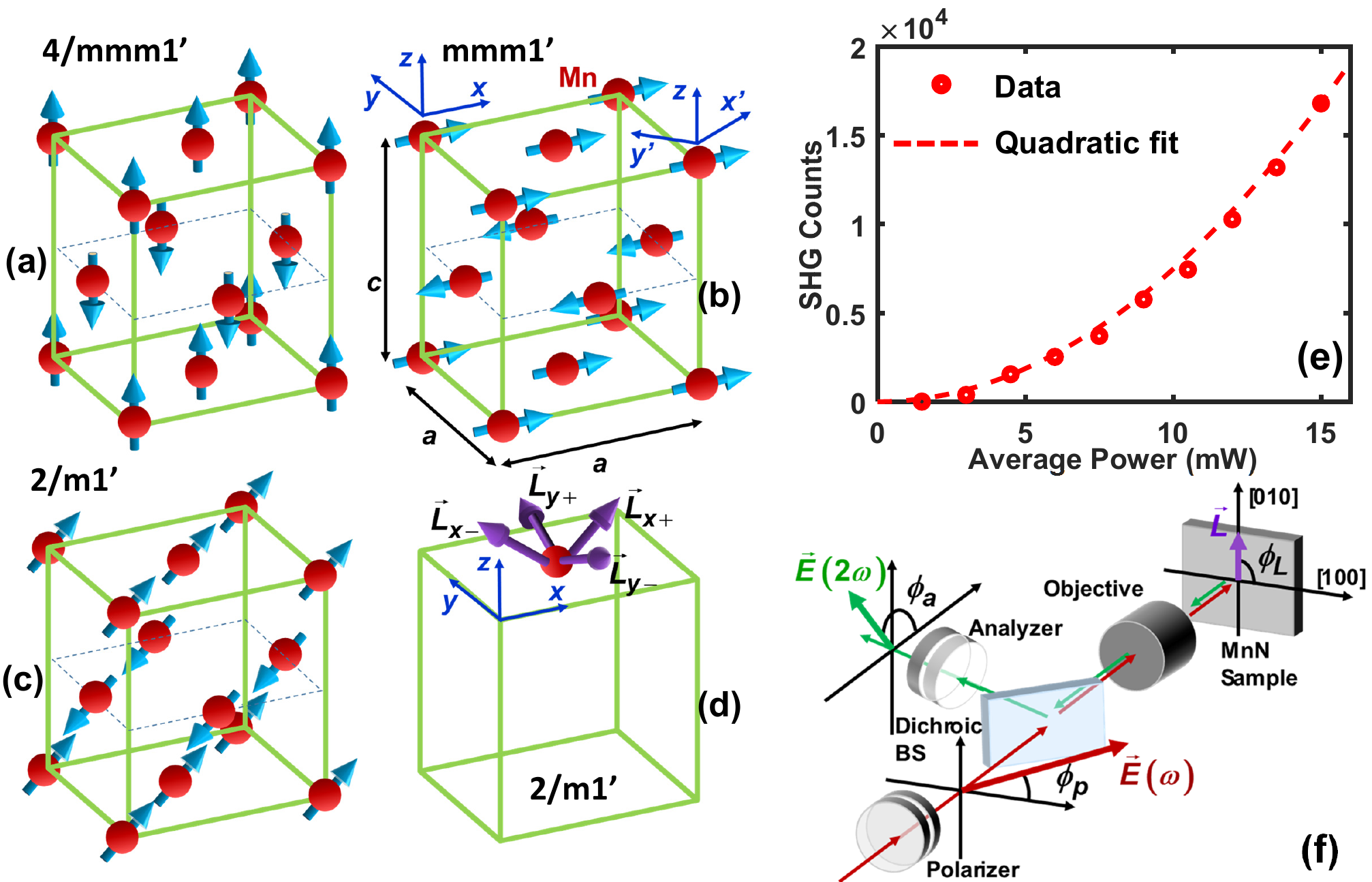}
\caption{\label{mpg} MnN unit cell with different spin structures corresponding to (a) 4/mmm1', (b) mmm1', and (c) 2/m1' magnetic points group symmetries are shown. (d) Neel vectors $\vec{L}$ for different magnetic domains consistent with 2/m1' symmetry are depicted. (e) Measured SH signal from a MnN sample showing a quadratic dependence on the pump power (T=300 K). (f) Experimental setup for measuring SHG is shown. $\phi_{p}$, $\phi_{a}$, and $\phi_{L}$ are the angles of the polarizer, analyzer, and the in-plane component of the Neel vector $\vec{L}$ with respect to the crystal [100] axis.}
\end{figure}

$\theta$-phase MnN has a tetragonally distorted rock-salt crystal structure. The Mn spin moments within a (001) plane are known to be aligned ferromagnetically, and they are antiferromagnetically aligned in different (001) planes~\cite{Suzuki2001, Leineweber2000,Kleinman2003,Lamb2003,Lamb2005,Zunger2008}(see Fig.~\ref{Growth}(a)). The exact orientation of the spin moments (and the magnetic point group) remain uncertain. Neutron diffraction studies have yielded conflicting results. Spins pointing within the (001) plane (mmm1' magnetic point group symmetry) were reported by K. Suzuki et al.~\cite{Suzuki2001}, and spins pointing in a direction tilted 67° out of the (001) plane (2/m1' symmetry) were reported by Leineweber et al.~\cite{Leineweber2000}. In addition, spins pointing along the [001] direction (4/mmm1' symmetry) have been predicted theoretically~\cite{Hong2005} and were also reported in measurements done at high temperatures~\cite{Leineweber2000}. MnN unit cells with different magnetic point group symmetries are depicted in Fig.~\ref{mpg}(a-c). The four Neel vectors $\vec{L}$ corresponding to the four different magnetic domains consistent with the 2/m1' symmetry are shown in Fig.~\ref{mpg}(d). Note that the magnetic point group symmetry remains unchanged if the components of the spin moments in the (001) plane are aligned along either one of the two sets of symmetry axes, <100> and <110>. In specifying the magnetic point groups, we have used the convention that in an antiferromagnetically ordered periodic crystal the magnetic point group consists of all crystal point group operations, together with the time-inversion operation, that leaves the crystal (including the magnetic order) invariant to within a translation~\cite{Ron2005}. Mn:N ratios in the 6:5.85 to 6:5.95 range have been reported for $\theta$-MnN in the literature~\cite{Leineweber2000}. However, no long range order or structure has been found corresponding to the missing nitrogen atoms~\cite{Leineweber2000}. The nitrogen vacancies are thought to stabilize the rock-salt structure over the competing zinc-blende structure~\cite{Zunger2008}.      

\begin{figure}[t]
\includegraphics[width=0.8\columnwidth]{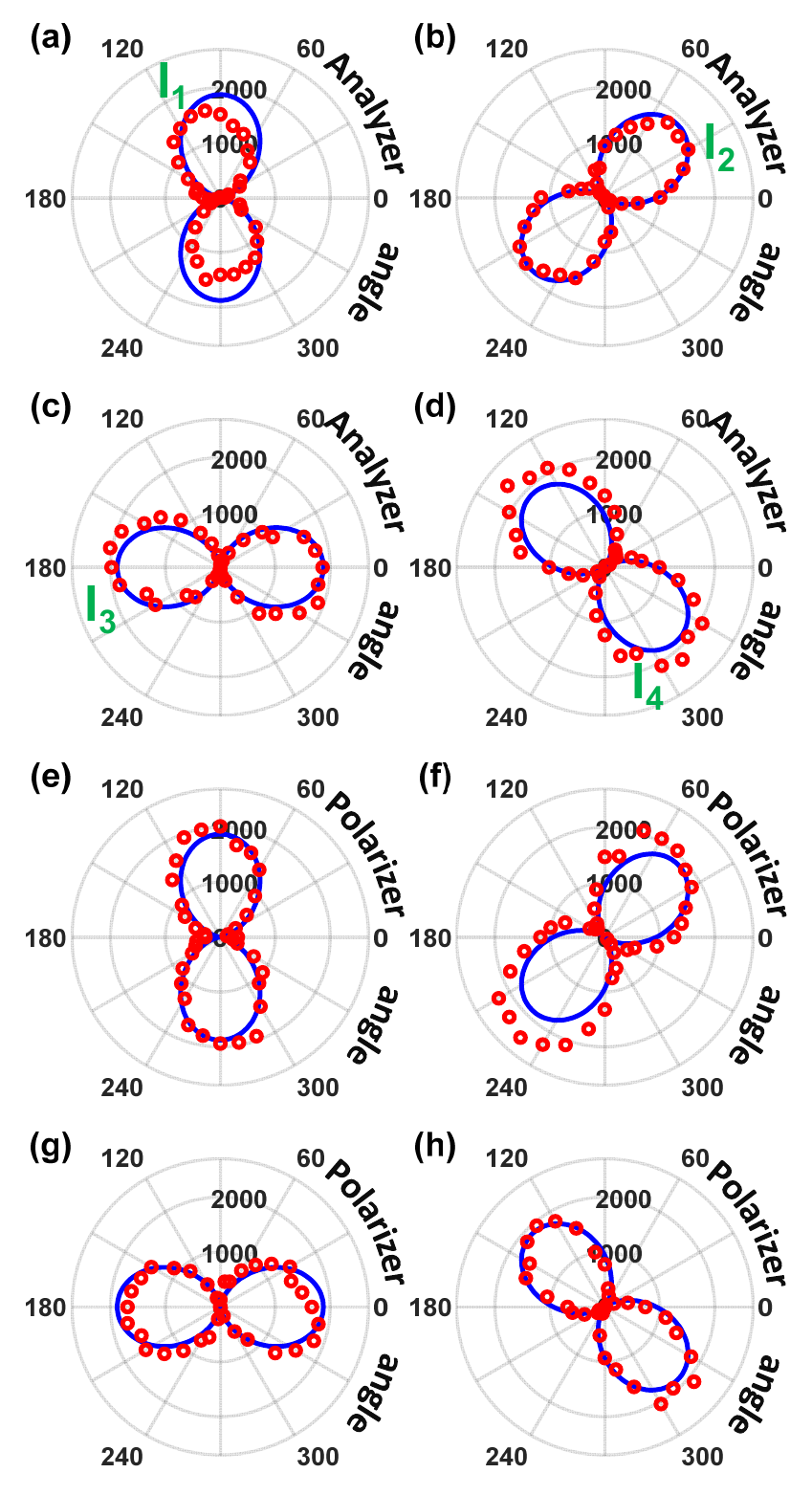}
\caption{\label{SHGRes1} SH photon counts are plotted as a function of the analyzer angle while the polarizer angle was fixed at (a) 0°, (b) -45°, (c) -90°, (d) -135°. Also shown are the SH counts as a function of the polarizer angle ($\alpha$) while the analyzer angle ($\beta$) was fixed at (e) 0°, (f) -45°, (g) -90°, (h) -135°. Experimental data from a single representative spot is shown is shown in red dots. Results from our theoretical model, discussed in the text, are also shown (solid lines). $I_{1}$, $I_{2}$, $I_{3}$, and $I_{4}$ are the peak intensities.}     
\end{figure}

\section{Material growth and characterization}

The MnN samples used in this study were grown using plasma-assisted molecular beam epitaxy (MBE) on (001) MgO substrates. High purity elemental Mn effusion cell supplied the metal flux, and ultra-high purity nitrogen gas was supplied through a plasma source. Growth was carried out at a thermal couple temperature of 425$\degree$C. Mn beam equivalent pressure (BEP) of $10^{-7}$ Torr and nitrogen plasma operating at 450 W with N$_{2}$ gas flow rate of 3.2 sccm was used for MnN layer deposition. Symmetric $2\theta/\omega$ X-ray diffraction (XRD) scan is shown in Fig.~\ref{Growth}(c). The observation of only integer orders of MnN (002) and MgO (002) peaks suggests that the MnN layers are single-crystal, with c-axis out-of-plane, and have no appreciable alien phase inclusion. Atomic force microscopy (AFM) measurements (Fig.~\ref{Growth}(b)) on a 500 nm thick MnN sample revealed a smooth surface with 0.6 nm root mean square (RMS) roughness. Resistivity of 90 $\mu\Omega$ cm, carrier concentration of 4$\times10^{21}$ cm$^{-3}$ and carrier mobility of 17 cm$^2$ V$^{-1}$ s$^{-1}$ obtained from Hall measurements revealed metallic properites. The sign of the charge of the charge carriers was found to be negative consistent with the computed electronic bandstructure of antiferromagnetic MnN~\cite{Lamb2003}. Vibrating sample magnetometry (VSM) with magnetic field applied in the out-of-plane direction exhibited no hysteresis loop up to 6 T (Fig.~\ref{Growth}(d)) ruling out inclusion of ferrimagnetic precipitates, such as Mn$_4$N, in the films. Nonlinear optical measurements (SHG) were performed on thinner MnN samples (in the 10-20 nm range) since most of the pump was completely absorbed in the first few tens of nanometers of the sample.  

\section{Optical second harmonic generation}

Second harmonic (SH) measurements were conducted using a 1032 nm center wavelength, $\sim$150 fs pulse width, Ytterbium laser with a $\sim$100 MHz repetition rate. All measurements were performed with 6 mW average power at room temperature. To minimize sample heating, the pump was mechanically chopped at 100 Hz with 30\% duty cycle. The pump was focused down to a $\sim$2.5 $\mu$m diameter optical spot on the sample at normal incidence. SH signal was collected in a reflection geometry. The quadratic dependence of the collected SH intensity on the pump power was confirmed (Fig.~\ref{mpg}(e)). The polarization angles of the incident pump (polarizer angle) and collected SH signal (analyzer angle) were selected using a set of half-wave plates and Glan Thompson polarizers (see Fig.~\ref{mpg}(f)). Fig.~\ref{SHGRes1} shows the SHG signal dependence on the polarizer and analyzer angles in eight different configurations. First, the polarizer was fixed at (a) 0°, (b) -45°, (c) -90°, (d) -135° while the analyzer was rotated through 360°. Second, the polarizer was rotated through 360° while the analyzer was fixed at (e) 0°, (f) -45°, (g) -90°, (h) -135°. All angles indicated are with respect to the [100] axis. The SH data was collected at a large number of points in a 100 $\mu$m$^{2}$ area grid on the samples. Two important features visible in the data in Fig.~\ref{SHGRes1} are, (i) SH measured is always polarized perpendicular to the pump, and (ii) peak SH intensity is independent of the pump polarization. In the Sections that follow, we will analyze these results, propose a model to explain the observed features, and relate the observations to the magnetic structure of the MnN films.   

\section{Theoretical modeling and discussion}

Although the centrosymmetric structure of MnN does not allow SHG (of pure electric dipole origin) in the bulk, broken inversion symmetry at the surface could generate a strong electric dipole SH signal that can be stronger than the bulk SH signal. We first note that the (001) surface of MnN has an in-plane $C_{4v}$ symmetry (in the absence of any surface magnetic order) which does not allow surface-normal electric dipole SHG. To make the case stronger against the measured SH signal coming from the surface, and rule out the possibility of the surface magnetic order contributing to the measured SH signal, we characterized samples with different thicknesses. Thicker samples generated more SH signal than thinner samples. The dependence of the measured SH intensity on the sample thickness $d$ was fitted with the approximate relation $1-\exp(-2\alpha(\omega)d-\alpha(2\omega)d)$ with the combined fundamental and SH loss, $2\alpha(\omega)+\alpha(2\omega)$, equal to $\sim 7.5 \times 10^{7}$ 1/m. For example, a 20 nm thick sample generated on average $\sim$15\% more SH signal intensity than a 15 nm thick sample. We, therefore, rule out any significant surface contribution to the measured SH signals. The presence of large number of nitrogen vacancies in $\theta$-MnN, as reported previously~\cite{Leineweber2000}, can contribute to the breaking of local inversion symmetry in the bulk. However, given the lack of any long range order or structure in the positions of the missing nitrogen atoms~\cite{Leineweber2000}, inversion symmetry is expected to remain a good symmetry (on the scale of the optical wavelength) as far as SHG is concerned.     

We assume that the Neel vector in the bulk is $\vec{L} = \vec{m}_{1} - \vec{m}_{2}$, and its component in the (001) plane is $\vec{L}_{\parallel}$. SHG in the centrosymmetric bulk due to magnetic dipole and electric quadrupole transitions can be expressed as,
\begin{eqnarray}
P_{i}(2\omega) & = & \{ \chi^{eem}_{ijk} + \gamma^{eem}_{ijk}(\vec{L}) \} E_{j}(\omega)H_{k}(\omega) \nonumber \\
& & +  \{ \chi^{eee}_{ijkl} + \gamma^{eee}_{ijkl}(\vec{L}) \} E_{j}(\omega)\partial_{l} E_{k}(\omega)\label{shg_desc}.
\end{eqnarray}
Here, the first two terms describe a second order process in which one of the intermediate transitions is of magnetic dipole origin~\cite{Sanger2006} and the next two terms describe a process in which one of the intermediate transitions is of electric quadrupole origin. The rationale for selecting magnetic dipole and electric quadrupole intermediate transitions will be discussed later. When allowed, both magnetic dipole and electric quadrupole transitions can be of comparable strengths in transition metal ions and compounds~\cite{Griffith1971}. The components of the susceptibility tensors that do not vanish as a consequence of the crystal and magnetic symmetries can be determined as follows. $\chi^{eem}_{ijk}$ and $\chi^{eee}_{ijkz}$ do not depend on the magnetic order and their non-zero components are determined by the 4/mmm structural symmetry of the crystal. Both these tensors have no non-zero components that can contribute to SHG in the surface-normal direction. The non-zero components of $\gamma^{eem}_{ijk}$ and $\gamma^{eee}_{ijkz}$ are determined by the magnetic point group symmetry of the crystal. Assuming surface-normal direction of the pump, in the case of both 4/mmm1' and mmm1' magnetic point group symmetries (Fig.~\ref{mpg}(a,b)), $\gamma^{eem}_{ijk}$ and $\gamma^{eee}_{ijkz}$ have no non-zero components that can generate a SH signal in the surface normal direction. If the magnetic point group symmetry is 2/m1' (Fig.~\ref{mpg}(c)), both $\gamma^{eem}_{ijk}$ and $\gamma^{eee}_{ijkz}$ have four non-zero components that can contribute to a SH signal in the surface-normal direction. Assuming $\vec{L}_{\parallel}$ is along the [100] direction, these components are: (i) $\gamma^{eem}_{yxx}$, $\gamma^{eem}_{xyx}$, $\gamma^{eem}_{xxy}$, $\gamma^{eem}_{yyy}$ and, (ii) $\gamma^{eee}_{yxyz}$, $\gamma^{eee}_{xyyz}$, $\gamma^{eee}_{xxxz}$, $\gamma^{eem}_{yyxz}$. If $\vec{L}_{\parallel}$ is along any other direction (e.g. [110] or [010]), the non-zero tensor components can be obtained by a rotation of the coordinates~\cite{Birss1966}. The analysis here shows that non-zero susceptibility tensor components that contribute to the SH signal in the surface normal direction result from the reduced symmetry generated by the magnetic order and are therefore a good probe of the magnetic order. 

We should point out that symmetry in many cases allows non-zero tensor components that generate a surface-normal polarization $P_{z}(2\omega)$. Although $P_{z}(2\omega)$ will not contribute to SH in the strictly surface-normal direction, it can contribute to SH at non-zero angles from the surface-normal. The amount of SH signal generated by $P_{z}(2\omega)$ that is collected in a surface-normal reflection-based measurement setup, such as ours, is determined by the numerical aperture (NA) of the objective. SH generated by $P_{z}(2\omega)$ itself will exhibit intensity that is independent of the analyzer angle. Interference with other in-plane polarization components could result in asymmetric lobes in the patterns displayed in Fig.~\ref{SHGRes1}. However, in our experiments use of collection objectives with different numerical apertures (NAs in the 0.1-0.6 range) resulted in no discernible changes in the patterns displayed in Fig.~\ref{SHGRes1}. Also note that in the case of 4/mmm1' and mmm1' magnetic point group symmetries, any contribution to the SH signal from $P_{z}(2\omega)$ would have resulted in the SH intensity being independent of the analyzer angle, in disagreement with the measured results in Fig.~\ref{SHGRes1}. Next, we show that the non-zero components of $\gamma^{eem}_{ijk}$ and $\gamma^{eee}_{ijkz}$, assuming 2/m1' magnetic point group symmetry, can explain all our SH data provided a statistical mixture of domains is assumed within the pump spot size.      

We first focus on the tensor $\gamma^{eem}_{ijk}$ and assume that $\vec{L}_{\parallel}$ is along the [100] axis. Since the measured SH is always perpendicular to the pump, we require that $\gamma^{eem}_{xxy}=\gamma^{eem}_{yyy}=0$ and $\gamma^{eem}_{yxx}= - \gamma^{eem}_{xyx}$. We are then left with only one independent component for the $\gamma^{eem}_{ijk}$ tensor which can contribute to surface-normal SHG. Assuming a single magnetic domain, if the angles with respect to the [100] axis of $\vec{L}_{\parallel}$, the analyzer, and the polarizer are $\phi_{L}$, $\phi_{a}$, and $\phi_{p}$, respectively, then the SH intensity in the surface-normal direction comes out to be proportional to,
\begin{equation}
  \sin^{2}(\phi_{a} - \phi_{p}) \sin^{2}(\phi_{L} - \phi_{p}).
\end{equation}
The SH signal obtained above exhibits a single-lobe polarized perpendicular to the pump and the SH intensity goes to zero when the pump is polarized parallel to $\vec{L}_{\parallel}$. In principle, SHG can therefore be used to detect the orientation of the in-plane component $\vec{L}_{\parallel}$ of the Neel vector provided the sample consists of a single domain. In experiments, the SH signal is always found to be polarized perpendicular to the pump consistent with the model (see Fig.~\ref{SHGRes1}(a-d)). However, the peak SH intensity is seen to be independent of the pump polarization. The experimental results can be explained by assuming many magnetic domains within the pump optical spot size of $\sim$2.5 $\mu$m. The Neel vectors of the four possible magnetic domains consistent with 2/m1' magnetic point group symmetry are shown in Fig.~\ref{mpg}(d) and correspond to $\phi_{L}$ values of $0$, $\pm \pi/2$, and $\pi$. We assume, for simplicity, that the sizes of individual domains are much smaller than the pump spot size and the SH wavelength. This assumption will be relaxed later when we present results from the computational model. We assume that the pump is polarized at an angle $\phi_{p}$ with respect to the sample [100] axis. The far-field SH intensity $I_{SH}(\phi_{a},\phi_{p},\theta, \phi)$ in the $(\theta, \phi)$ direction, collected by the objective, and after passing through the analyzer fixed at angle $\phi_{a}$, can be obtained by coherently superposing SH field emitted by all the domains within the pump spot size. The spherical coordinate angles $\theta$ ($\phi$) are measured with respect to the [001] ([100]) axis. The result, for small values of $\theta$, can be written as,
\begin{eqnarray}
I_{SH}(\phi_{a},\phi_{p},\theta, \phi) & \propto & \sin^{2}(\phi_{a}-\phi_{p})
\sum_{m,n} \sin(\phi^{m}_{L}-\phi_{p}) \sin(\phi^{n}_{L}-\phi_{p})\nonumber \\
& & \times e^{-ik \sin\theta [ (x^{m}-x^{n})\cos\phi + (y^{m}-y^{n})\sin\phi ]}.
\end{eqnarray}
The sum over $m/n$ is over all the $N$ domains located at coordinates $(x^{m/n},y^{m/n})$ in the sample within the pump spot size. If the domains are much smaller than the pump spot size and the SH wavelength, the terms in this sum interfere with each other strongly, and the sum is then dominated by the diagonal terms for which $m=n$. Keeping only these terms gives the approximate expression,
\begin{eqnarray}
I_{SH}(\phi_{a},\phi_{p},\theta, \phi) & \propto & \sin^{2}(\phi_{a}-\phi_{p}) \sum_{m} \sin^{2}(\phi^{m}_{L}-\phi_{p}) \nonumber \\
  & \approx &  \frac{N}{2} \sin^{2}(\phi_{a}-\phi_{p}) \label{pol_result_1}.
\end{eqnarray}
The above result corresponds to a single-lobe SH signal polarized perpendicular to the pump and whose peak intensity is independent of the pump polarization. The expression in Eq.(\ref{pol_result_1}), plotted as solid line in Fig.~\ref{SHGRes1}(a-h), is seen to match the SH data very well for all polarizer and analyzer angles. Note that if $\vec{L}_{\parallel}$ is assumed to be along the <110> directions (which is also consistent with a 2/m1' magnetic point group), and $\phi_{L}$ has values $\pm \pi/4$ and $\pm 3\pi/4$, the result in Eq.(\ref{pol_result_1}) will remain unchanged. Therefore, if the magnetic domains are much smaller than pump spot size and the SH wavelength, our measurements cannot distinguish between <100> and <110> orientations of $\vec{L}_{\parallel}$.  

\begin{figure}[t]
\includegraphics[width=1.0\columnwidth]{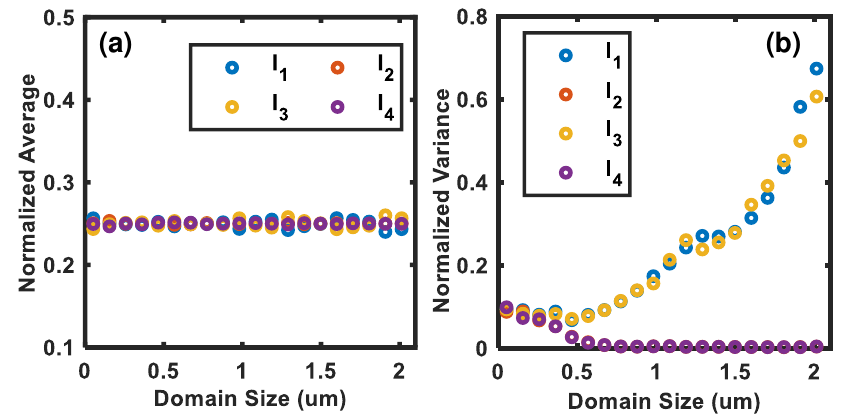}
\caption{\label{EnsStats} Calculated normalized average (a) and normalized variance (b) of the peak SH intensities $I_1$, $I_2$, $I_3$, and $I_4$ (for different pump polarization, see Fig.~\ref{SHGRes1}) are plotted as a function of the magnetic domain size.}
\end{figure}

We make two remarks about the analysis presented above. First, in actual experiments SH signal is collected within a cone determined by the collection objective's NA and the values of $\theta$ for which the SH signal is recorded are not always very small. Second, suppose the domain sizes increase from very small values to sizes comparable to the pump spot size and the SH optical wavelength ($\sim$0.5 $\mu$m). One then ought to observe larger fluctuations in the measured SH intensities as one collects SH signal from different spatial spots on the sample. This is noteworthy because the strength of the measured fluctuations in the SH data collected from a large number of spots on the sample can be used to place an upper limit on the domain sizes in the sample. To model this phenomenon more carefully, we developed a finite-element computational model in which we divided the sample area within the pump spot size into different square-shaped magnetic domains of a given size, and then divided each domain further into smaller pixels of size ~10 nm. We then computed the SH far-field from the bulk SH polarization $P(2\omega)$ in every pixel. The SH field within the collection light cone (cone angle determined by the objective's NA assumed to be 0.5) was collimated and squared to obtain the SH intensity. SH intensity thus obtained was ensemble averaged. A typical ensemble consisted of $\sim$500 different magnetic domain configuration within the pump spot size, corresponding to $\phi_{L}$ values of $0$, $\pm \pi/2$, and $\pi$. The results obtained are shown in Fig.~\ref{EnsStats} which plots the normalized average (a) and the normalized variance (b) of the peak SH intensities $I_1$, $I_2$, $I_3$, and $I_4$ (for different pump polarizations, see Fig.~\ref{SHGRes1}) as a function of the magnetic domain size. The normalized average $e[I_{j}]$ and variance $v[I_{j}]$ are defined as, $e[I_j]=E[I_j]/\sum_{j} E[I_j]$ and $v[I_j]=(E[I_j^2]-E[I_j]^2)/E[I_j]^2$. In the limit of very large domain sizes, one expects $v[I_{2/4}]$ to go to zero and $v[I_{1/3}]$ to approach unity  given the two non-zero components of the susceptibility tensor $\gamma^{eem}_{yxx}= - \gamma^{eem}_{xyx}$. The computation shows that the fluctuations in $I_{1/3}$ become large once the domain sizes exceed the SH wavelength ($\sim 0.5$ $\mu$m). In experiments, ensemble averaging was performed by recording the SH signal from different spots on the sample. The experimental value of $e[I_{j}]$ was $\sim 0.25$ for all $I_{j}$, in very good agreement with the computational model. The experimental values of $v[I_{j}]$ were found to be smaller than $.02$ for all $I_{j}$. This value is slightly smaller than the smallest computed values for $v[I_{2/4}]$ of $\sim.04$ and for $v[I_{1/3}]$ of $\sim.07$ which occur for a domain size of $\sim0.45$ $\mu$m. This small discrepancy can be attributed to our use of the same domain shape and size in all ensembles. A more complex model that uses domains of slightly different shapes and sizes within the same ensemble would be closer to reality and is expected to reduce the computed variance in the limit of small domain sizes. The comparison between experiments and the model enables us to place a rough upper bound on the domain sizes in our samples of $\sim 0.65$ $\mu$m given that the variances $v[I_{1/3}]$ become large for larger domain sizes. This procedure does not provide a tight upper bound.                         

\begin{figure}[t]
\includegraphics[width=1.0\columnwidth]{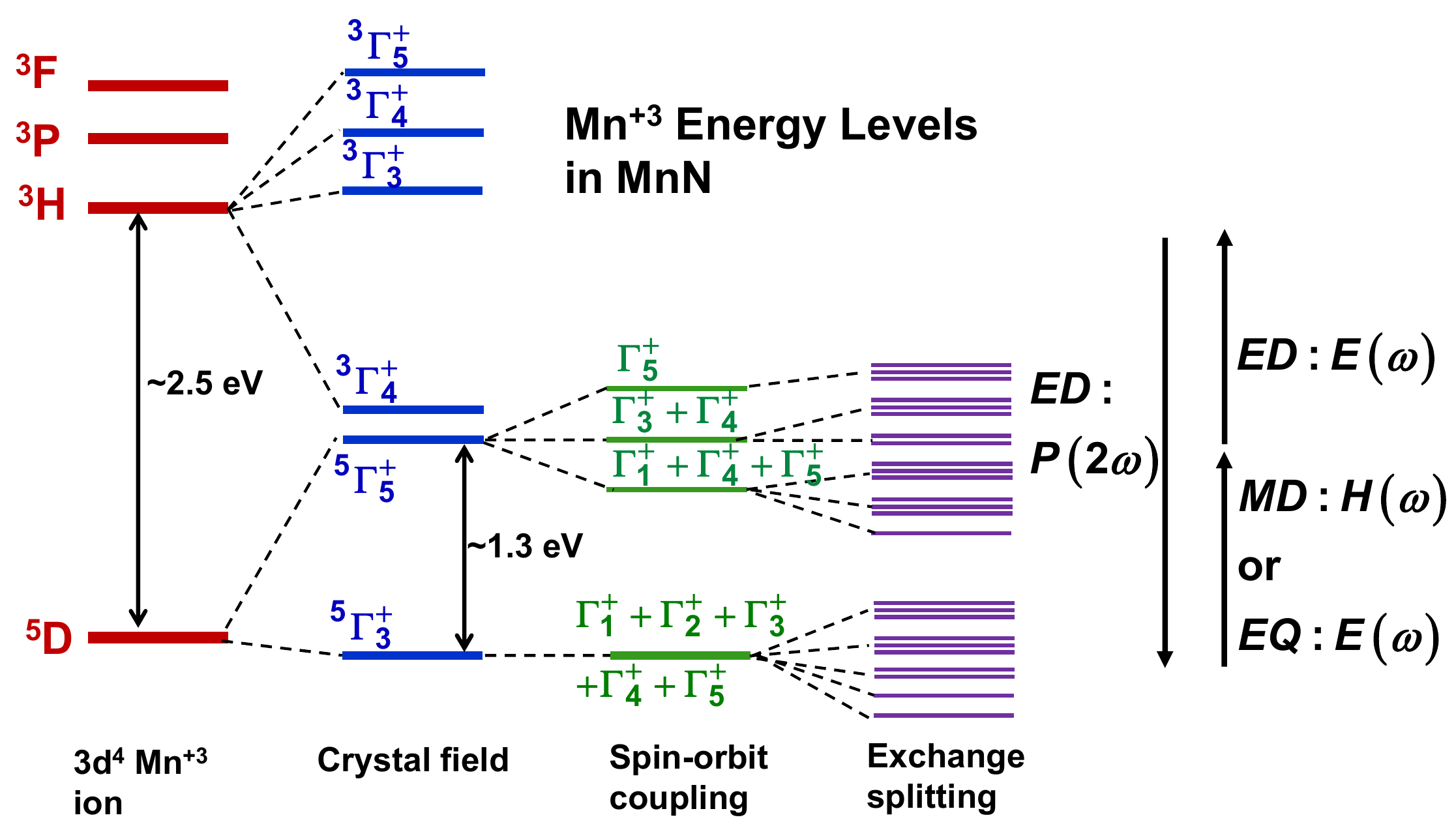}
\caption{\label{Mn} Energy levels of Mn$^{3+}$ in MnN are shown~\cite{Konig1974,Figgis1999}. Both magnetic dipole and electric quadrupole transitions are allowed between the lowest energy $^{5}\Gamma_{3}$ multiplets and the multiplets of the upper $^{5}\Gamma_{5}$  level. Energy level splittings due to the small tetragonal distortion in MnN have been ignored. The energy level splittings shown are not drawn to scale.}    
\end{figure}

Next, we focus on the non-zero components of the tensor $\gamma^{eee}_{ijkz}$ that results from an electric quadrupole intermediate transition. The symmetry properties of $\gamma^{eee}_{ijkz}$ are very similar to those of $\gamma^{eem}_{ijk}$ and, quite remarkably, the assumptions $\gamma^{eee}_{xxxz}=\gamma^{eee}_{yyxz}=0$ and $\gamma^{eee}_{yxyz}= - \gamma^{eee}_{xyyz}$ can explain our data just as well as the tensor $\gamma^{eem}_{ijk}$ discussed above. To explore the origin of the SHG in our samples, and possibly single out one tensor responsible for the SHG, we look at the electron energy levels of d-orbitals in Mn$^{3+}$ (3d$^{4}$) in the MnN environment~\cite{Konig1974,Figgis1999}.

\section{Optical transitions and the origin of SH polarization}

The bands near the Fermi energy in MnN are formed by weakly coupled d-orbitals of adjacent Mn atoms. The nitrogen p-orbitals in MnN have been found to be lower in energy~\cite{Kleinman2003,Zunger2008}. It is therefore reasonable to assume that the selection rules for optical transitions can be determined from the symmetries of the energy levels of a Mn$^{3+}$ ion in the tetragonal 4/mmm environment. Since the tetragonal distortion in MnN is small (Mn-N bond length is 2.128 {\AA} (2.095 {\AA}) along the a-axis (c-axis)~\cite{Lamb2005}), the local environment of each Mn atom can be approximated by the octahedral ($O_{h}$) group rather than the $D_{4h}$ group without affecting the main results that follow. The lowest energy levels, and their symmetries, of Mn$^{3+}$ in an octahedral environment are depicted Fig.~\ref{Mn}. The lowest occupied level has $^{5}\Gamma^{+}_{3}$ symmetry and in an octahedral environment $^{5}\Gamma^{+}_{3}=\sum^{5}_{i=1} \oplus \Gamma^{+}_{i}$. The magnetic dipole transition operator has the symmetry $\Gamma^{+}_{4}$, and since $^{5}\Gamma^{+}_{3} \otimes \Gamma^{+}_{4} = \Gamma^{+}_{1} \oplus \Gamma^{+}_{2} \oplus 2\Gamma^{+}_{3} \oplus 4(\Gamma^{+}_{4} \oplus \Gamma^{+}_{5})$, magnetic dipole transitions are allowed between $^{5}\Gamma^{+}_{3}$ and all multiplets of the higher energy $^{5}\Gamma^{+}_{5}$ level. The electric quadrupole transition operator has the symmetries $\Gamma^{+}_{5}$ (off-diagonal components) and $\Gamma^{+}_{3}$ (diagonal components). And since $^{5}\Gamma^{+}_{3} \otimes \Gamma^{+}_{5} = \Gamma^{+}_{1} \oplus \Gamma^{+}_{2} \oplus 2\Gamma^{+}_{3} \oplus 4(\Gamma^{+}_{4} \oplus \Gamma^{+}_{5})$ and $^{5}\Gamma^{+}_{3} \otimes \Gamma^{+}_{3} = \Gamma^{+}_{1} \oplus \Gamma^{+}_{2} \oplus 3\Gamma^{+}_{3} \oplus 2(\Gamma^{+}_{4} \oplus \Gamma^{+}_{5})$, electric quadrupole transitions are also allowed between $^{5}\Gamma^{+}_{3}$ and all multiplets of the higher energy $^{5}\Gamma^{+}_{5}$ level. Two possibilities behind the origin of SHG in MnN are depicted by the transition diagram in Fig.~\ref{Mn} which shows that one intermediate transition, involving the ground state, of the second order nonlinear process is either of magnetic dipole or of electric quadrupole origin, and the other two transitions, involving higher energy states, are of electric-dipole origin. The rationale for assigning transitions in this way is as follows. The transition from $^{5}\Gamma^{+}_{3}$ to $^{3}\Gamma^{+}_{4}$, and also transitions from $^{5}\Gamma^{+}_{3}$ and $^{5}\Gamma^{+}_{5}$ to other higher energy levels, require a spin-flip and are expected to be weak. Since even spin-allowed magnetic dipole and electric quadrupole transitions are weak (much weaker than electric-dipole transitions)~\cite{Griffith1971}, we expect the $^{5}\Gamma^{+}_{3}$ to $^{5}\Gamma^{+}_{5}$ spin-allowed transition to be of magnetic dipole or electric quadrupole character. Electric dipole transitions involving higher energy levels, although spin and dipole forbidden at the level of a single Mn ion, are expected to be strong and made possible by the larger coupling between d-orbitals of adjacent Mn atoms at higher energies, enhanced $d-p$ coupling at higher energies, and the presence of spin-orbit coupling~\cite{Figgis1999, Griffith1971, Kleinman2003}. Finally, since the pump and SHG photon energies in our experiments are $\sim$1.2 eV and $\sim$2.4 eV, respectively, we don't expect any one of the transitions contributing to SHG in our experiments to be fully resonant. 

\section{Conclusion}

In conclusion, SHG measurement scheme is found to be sensitive to the antiferromagnetic order in MnN and enabled the selection of 2/m1' as the magnetic point group symmetry of MnN from among other competing candidates. Our work shows that SHG can be used to place a loose upper bound on the domain sizes. Our work also shows that SHG can be used to probe the magnetic order in bulk metallic antiferromagnets. Previously, this technique has been used to probe the magnetic order in bulk insulating antiferromagnets~\cite{Fiebig2005,Sanger2006}. We expect that in single-domain MnN samples this technique can be used to detect the in-plane orientation of the Neel vector and also its switching in response to stimuli such as the spin transfer torque~\cite{Rev1,Rev2,Dunz2020}. Nonlinear optical techniques can thus play an important role in antiferromagnetic spintronics. 

\section*{Author Contributions}
J.L. and Z.Z. contributed equally to this work.

\begin{acknowledgments}
This work was supported by the Cornell Center for Materials Research with funding from the NSF MRSEC program (DMR-1719875). The authors would like to acknowledge helpful discussions with Okan Koksal, Gregory D. Fuchs, and Daniel C. Ralph. The authors would like to thank Joseph Casamento for x-ray photoelectron spectroscopy measurements. 
\end{acknowledgments}

\section*{Appendix: Temperature Dependence of the Second Harmonic Intensity}
\begin{figure}[t]
\includegraphics[width=0.7\columnwidth]{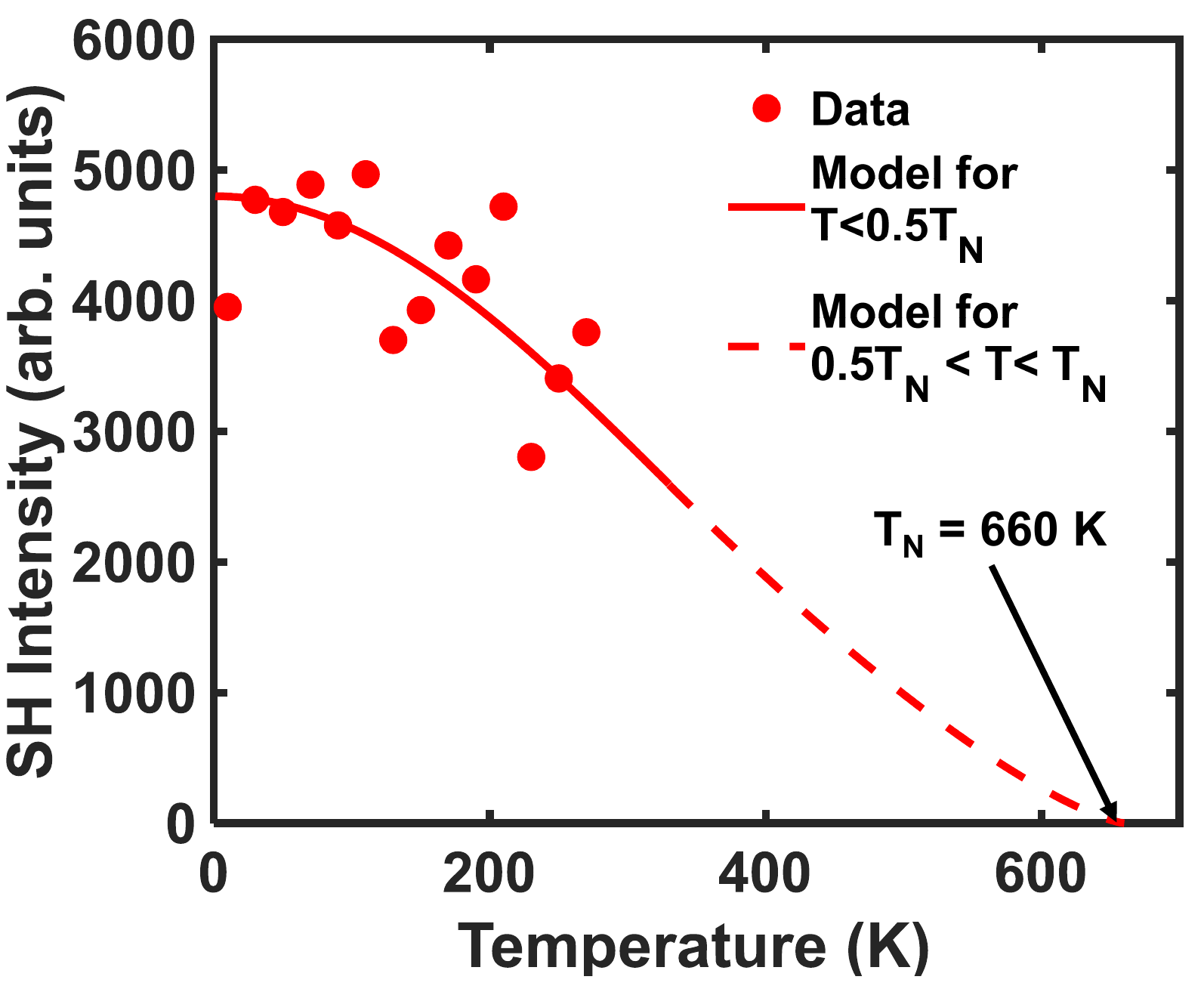}
\caption{\label{TDep} Measured SH intensity (for polarizer and analyzer angles that yield the maximum SH intensity) for MnN is plotted as a function of the temperature $T$ (circles). The solid and dashed lines show the fit obtained using the models discussed in the text valid for $0 < T < 0.5 T_{N}$ and $0.5T_{N} < T < T_{N}$, respectively.}    
\end{figure}

In the discussion following Eq.~\ref{shg_desc}, we argued that the only contributions to the measured SH intensity in the surface-normal direction can come from the components $\gamma^{eem}_{ijk}(\vec{L})$ and $\gamma^{eee}_{ijkl}(\vec{L})$ of the susceptibility tensor that explicitly depend on the Neel vector $\vec{L}$. Since in antiferromagnets magnitudes of the Neel vector components are temperature dependent~\cite{Kittel1963}, one should expect the measured SHG intensity to be temperature dependent as well. Fig.~\ref{TDep} shows the  measured SH intensity as a function of the temperature. The small spread in the data is attributed to thermal expansions/contractions in the sample mount inside the cryostat with temperature changes that are expected to result in a shift of the optical focus spot by tens of nanometers, and perhaps a few hundred nanometers, which are comparable to the domain sizes in the sample. The data shows that the SH intensity decreases as the temperature increases. Due to the limitations of our experimental setup, data beyond room temperature could not be obtained.

To explain the data, we assume the simplest model that the decrease in the sublattice magnetization is caused by excitation of the antiferromagnetic magnons~\cite{Kittel1963}. If the magnons have a spin gap $\Delta$, as is often the case in antiferromagnets with anisotropy~\cite{Kittel1963}, then for temperatures lower than $\sim$$\Delta/(3k_{B})$, the temperature dependence of the Neel vector is given as~\cite{Rado1965},
\begin{equation}
  L_{j}(T) \propto \left[ 1 - Ae^{-\Delta/k_{B}T} \left(\frac{T}{T_{N}}\right)^{3/2}\right]
\end{equation}
where $A$ is a constant, and for temperatures higher than $\sim$$\Delta/(3k_{B})$, but smaller than $\sim$$0.5T_{N}$, the temperature dependence of the Neel vector is described by~\cite{Kittel1963,Rado1965},
\begin{equation}
  L_{j}(T) \propto \left[ 1 - B\left(\frac{T}{T_{N}}\right)^{2}\right] \label{eq:LT}
\end{equation}
Here, $B\approx 0.57$. Since the spin gap $\Delta$ is not larger than a few meV in almost all metallic antiferromagnets, we expect that the expression in Eq.~\ref{eq:LT} to better describe the temperature dependence of our data for $10 K < T < 0.5T_{N}$. For even higher temperatures, $0.5 T_{N} < T < T_{N}$, the temperature dependence of $L_{j}(T)$ has been shown to be given by~\cite{Vanl1966},
\begin{equation}
L_{j}(T) \propto \left[ 1 - \frac{T}{T_{N}} \right]^{\beta}
\end{equation}
where the critical exponent $\beta$ is close to 1/3~\cite{Vanl1966}. To relate the temperature dependence of $\vec{L}(T)$ to  the temperature dependence of the SH intensity, we proceed as follows. The tensors, $\gamma^{eem}_{ijk}(\vec{L})$ and $\gamma^{eee}_{ijkl}(\vec{L})$, can both be expanded as follows,
\begin{equation}
\gamma_{ijk}(\vec{L}) = \gamma^{(1)}_{ijkr}L_{r} + \gamma^{(2)}_{ijkrs}L_{r}L_{s} + \ldots \label{eq:expansion}
\end{equation}
Since $\vec{L} \rightarrow - \vec{L}$ is a symmetry of the MnN crystal (up to a lattice translation), the first term in the expansion above that will contribute to the SH intensity in MnN is the second term with a quadratic dependence on the Neel vector components. Ignoring $\Delta$ for simplicity (since $\Delta$ is expected to be very small), the temperature dependence of the SH intensity $I(T)$ in MnN, for $T < 0.5T_{N}$ ought to be,
\begin{equation}
I(T) \approx I(T=0)\left[ 1-B\left(\frac{T}{T_N}\right)^2 \right]^{4} \label{eq:ISHG}
\end{equation}
The fit to our data obtained by using the expression above is shown by the solid line in Fig.~\ref{TDep}. The fit assumes a Neel temperature $T_{N}$ of 660 K. The data is seen to match the model very well for temperatures below the room temperature. The decay of the SHG intensity with temperature as $T^{2}$ is expected to be valid for $T<<T_{N}$ and our data is in agreement with this behavior.   

\bibliography{MnN_Reference}

\end{document}